\documentclass[twocolumn,pre,floatfix,showpacs]{revtex4}
\usepackage{dcolumn}
\usepackage{graphicx}
\usepackage{graphicx,amssymb,amsmath}
\usepackage{natbib}
\newcommand{\bra}[1]{\langle #1|}
\newcommand{\ket}[1]{|#1\rangle}

\begin{document}

%
%
%

\title{Intermolecular adhesion in conducting polymers}
\author{Jeremy D.\ Schmit}
\affiliation{Biomolecular Science and Engineering Program, University of California Santa Barbara,
CA 93106}\email[]{schmit@mrl.ucsb.edu}

\author{Alex J.~Levine}
\affiliation{Department of Physics, University of Massachusetts, Amherst MA 01060 USA.}
\email[]{levine@physics.umass.edu}

\begin{abstract}
We analyze the interaction of two conducting, charged polymer chains in solution using a minimal model for their electronic degrees of freedom. We show that a crossing of the two chains in which the polymers
pass within Angstroms of each other leads to a decrease of the electronic
energy of the combined system that is significantly larger than the thermal energy and thus promotes interchain aggregation. We consider the competition of this attractive interaction with the screened electrostatic repulsion and thereby propose a phase diagram for such polymers in solution; depending on the charge density and persistence length of the chains, the polymers may be unbound, bound in loose,
braid-like structures, or tightly bound in a parallel configuration.
\end{abstract}
\pacs{82.35.Cd,36.20.Kd,72.80.Le}

\maketitle

\section{Introduction}

Conjugated polymers have been the subject of intense research in the physics,
chemistry, and materials science communities. These molecules provide both a
laboratory to explore one-dimensional conductors \cite{Su:79,Su:80,Takayama:80}
and the potential for a plethora of applications in such disparate devices as
solar cells \cite{solar} and chemo/biosensors\cite{Heeger:98,Heeger:01,McQuade:00,Gaylord:03}.
The simplest conducting polymer system is polyacetylene. The underlying cause of its conductivity,
however, is somewhat subtle owing to the fact that, due to the Peierls
instability\cite{Peierls:55,Yannoni:83} in one dimension, this molecule
appears to be a semiconductor (based on an analysis of uncorrelated electrons).
Here we consider the case of an electron-doped polymer in which the Fermi energy lies within the conduction band.

The utility of conjugated polymers is generally realized in such complex chemical
systems created by adding functional groups to the basic, conjugated backbone.
Such modified polymers typically include side groups to modify the conductivity
of the backbone by doping in either electrons or holes; polar or ionizable side groups can also dramatically increase the solubility of these molecules in water \cite{Shi:90}.  Fluorescent, water-solubilized conjugated polymers such as polyparaphenylene vinylene (PPV) hold promise for use as biosensors since their fluorescence is readily quenched by electron acceptors such as methyl viologen (MV).  Biomolecules tagged with MV can be detected optically at remarkably low concentration using these water soluble conjugated polymers.  Fluorescence quenching can be quantified by the Stern-Volmer constant, $K_{\rm SV}$, which describes the loss of quantum efficiency per mole of quencher present. Since the $K_{SV}\simeq 10^7 M^{-1}$ for the PPV/MV system, MV tagged biomolecules can be detected via fluorescence quenching at 100 nanomolar concentrations\cite{Chen:99}.  Before this analytic tool can be realized, a number of complications must be addressed. These include understanding and accounting for the dramatic effect of ``bystander'' molecules such as surfactants on the value of $K_{SV}$\cite{Chen:00}, as well as addressing the role played by chain length of the PPV \cite{Gaylord:01,Wang:01}. Additionally, the PPV/MV system is hindered as an analytical tool by the tendency of the PPV to aggregate in solution and to bind nonspecifically with other small molecules.  Neutron and light scattering\cite{Wang:01a} demonstrate that PPV solutions form large aggregates under low-salt conditions. With added salt, the scattering data can be fit by rod-like structures having a persistence length of $80$nm \cite{Wang:01b} suggesting the bundling of the chains in solution.

Understanding the tendency of metallic polymers to aggregate in solution is the central focus of this paper. We propose an interchain adhesion mechanism specific to conjugated polymers that should apply to doped, solubilized PPV and other conducting polymers. When two polymer chains approach each other to Angstrom-scale distances at some point along their backbones, the interchain tunnelling of the electrons (holes) in the conduction bands of the two molecules decreases the total electronic energy of the system. This decrease in the energy of the electronic degrees of freedom is primarily due to the creation of a low energy, localized state at the crossing point of the two chains. These states lead to the aggregation of the two chains since, for physically reasonable parameters the energy decrease per crossing point is on the order of a few $k_{\rm B} T$.  Such a electron tunnelling mechanism has been previously discussed as a source of an {\em intrachain}, attractive interaction leading to polymer collapse in good solvent \cite{Hone:01}.

The mechanism that we propose can be thought of as analogous to the creation of a molecular covalent bond with the distinction that in the present system, the bonding and anti-bonding states that are localized at the crossing point of the two chains are not created from normal atomic energy eigenstates, but rather are pulled out of the delocalized states of the charge carriers on the chains.  In the remainder of the article, we first explore the consequences of this type of interchain bonding (in section \ref{adhesion}) in three stages of increasing complexity. We consider the effect of a single chain-crossing point that leads to interchain tunnelling in section \ref{single}. We then turn to the case of multiple crossing points by considering an ordered array of such points in section \ref{ordered}, before turning to the more physical case of a thermally disordered set of crossing points in \ref{disordered}.  Finally, we consider the attraction between two polymers aligned in a parallel configuration in \ref{railroad}.

In order to determine the adhesive properties of these conjugated polymers, we need to compare the decrease in the energy of the electronic degrees of freedom of the molecule, to the increase of chain conformational free energy associated with adhesion.  We turn to the latter calculation in section \ref{polymer}, before combining these two parts of the analysis in section \ref{conclusions} wherein we discuss the results and propose experimental tests of this work.

\section{Interchain Adhesion}
\label{adhesion}

Since we suggest that the adhesive properties of the conducting polymers are a generic consequence of populating
the conduction band of the these extended molecules, we
develop our theory based on a minimalistic, tight-binding Hamiltonian for these conduction electrons on a chain of
$N$ sites of the form
\begin{equation}
\label{tight-binding}
H_0=-t\sum_{\ell=-N/2}^{N/2} | \ell\rangle \langle \ell + 1 | + | \ell +1 \rangle \langle \ell |,
\end{equation}
where $|\ell\rangle$ is a state vector for an electron on the $\ell^{\rm th}$ tight binding site.
In the absence of strong electron--electron correlations\cite{Heeger:01} such a single particle
approach is justified and many properties of such conjugated polymers have been predicted via such simple tight-binding models \cite{Fink:86}. The overlap integral $t$ is not precisely known for a number of these systems, but recent spectroscopic results on chemically related systems\cite{Fink:86} suggest that the $t$ is on the order of $1 - 4 eV$.

\begin{figure}[htpb]
  \centering
  \includegraphics[width=8.0cm]{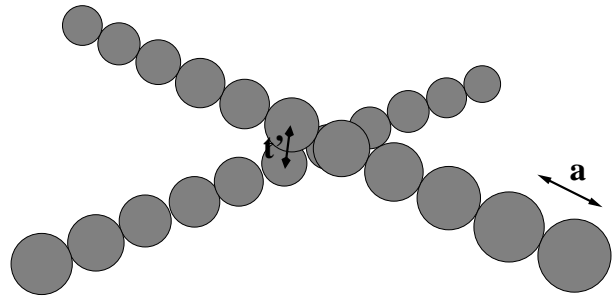}
  \caption{A pictorial representation of the simplest crossed--chains
  configuration in which the inter-chain hopping
  is allowed at the central site, $\ell = 0$. The spheres represent
  the tight-binding sites for the conduction electrons along the chains.
  The chains need not be straight as shown here.}
\label{model-setup}
\end{figure}

To discuss the modification of the electronic states of this tight-binding model due to the proximity of another such polymer at one point or a set of points $\left\{ \ell_I \right\}$, we modify Eq.~\ref{tight-binding} by introducing another copy and including an interaction term between the chains parameterized by a second hopping matrix element $t'$:
\begin{eqnarray}
\label{tb-interchain}
H_0 &=&  -t \sum_{j=1,2} \sum_{\ell=1}^N \left( |\ell + 1,j\rangle \langle \ell,j | + |\ell,j\rangle \langle \ell +1,j | \right) \\
\label{tb-interaction}
H_I &=& - t' \sum_{\bar{\ell} \in \left\{ \ell_I \right\} } \left( |\bar{\ell},1 \rangle \langle \bar{\ell},2|  + |\bar{\ell},2\rangle \langle \bar{\ell},1| \right).
\end{eqnarray}
In the above equations, the sum on $j$ is over the two chains, while the interaction Hamiltonian, $H_I$ allows for interchain hopping at a selected subset of sites along the polymer, $\left\{ \ell_I \right\}$. The interchain hopping matrix element has been shown numerically to be on the order of $0.1 eV$ for similar chemical systems\cite{Bredas:02}.  It is essential to note that since the interchain tunnelling matrix element is exponentially sensitive to the interchain separation, it is acceptable to suppose that this matrix element will be finite at only isolated points. In fact, we will later show in section \ref{polymer} that the positions of these ``crossing-points'' to be spatially correlated or anticorrelated along the arc length of the chain due to chain conformational free energy. A simple pictorial representation of the basic problem for the case of only one hopping site is shown in figure \ref{model-setup}.

Because of  the large separation of time scales between the
electronic and conformational degrees of freedom of the polymer
chain, we may ignore conformational changes in the polymer while
discussing the modification of the electronic states due to the
proximity of the two chains at a given point or set of points. To
show that we may ignore the thermal motion of the point of closest
approach in discussing the electronic structure of the molecules, we
compare the time scale for monomer motion over distances of a Bohr radius (the distance over which one expects $t'$ to
vary rapidly) $ T_{\rm chain} \sim a_{B} \sqrt{m/(k_{\rm B}
T)}$ to characteristic times for electronic structure
reorganization $T_{\rm electon} \sim \hbar/t$.  We find that
$T_{\rm electron}/T_{\rm chain} \sim 10^{-10}$ so the adiabatic
approximation is eminently reasonable.

Similarly, we note that the temperature of the system (typically
$300K$ or less) is significantly less than the Fermi temperature
of electronic system so in the remainder of the calculation we
will determine the total free energy of the electronic degrees of
the freedom in a zero temperature approximation whereas we will
consider the role of temperature and consequently entropy while
discussing the conformational and translational degrees of freedom of the polymer.
We now develop our calculation for the binding energy of the two
chains by computing separately the decrease in the energy of the
electronic degrees of freedom due to interchain tunnelling and
increase in chain free energy due to the loss of some
translational and conformational entropy. From the sum of these
two effects, we determine the effective binding energy. First we
consider a single crossing point as shown in figure
\ref{model-setup}.

\subsection{A Single Crossing Point}
\label{single}

\begin{figure}[htpb]
  \centering
  \includegraphics[width=8.0cm]{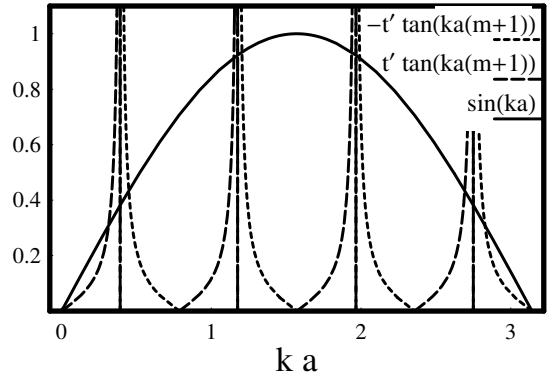}
  \caption{Graphical solution of Eq. \ref{quantizationII}.  Allowed $k$ values lie at the intersections of the solid line with the long dashes (symmetric band) and with the short dashes (anti-symmetric band).  The vertical lines represent the unperturbed $k$ values.}
\label{quant-fig}
\end{figure}

We take the single crossing point to lie in the middle of both chains $\{ \ell_I \} = \{ 0 \}$ as shown in figure \ref{model-setup}.  We return to crossing configurations of lower symmetry later. We note that the interaction Hamiltonian, Eq.~\ref{tb-interaction}, can be diagonalized in the basis of symmetrized and antisymmetrized chain occupation states:
\begin{equation}
\label{symm}
| \ell, \pm \rangle = \frac{1}{\sqrt{2}} \left[ | \ell, 1 \rangle \pm | \ell, 2 \rangle \right].
\end{equation}
so that in this basis Eq.~\ref{tb-interaction} takes the form
\begin{equation}
\label{diag-interchain}
H_I = -t' \sum_{\bar{\ell} \in \left\{ \ell_I \right\} } \left[ \, \ket{\bar{\ell},+} \bra{\bar{\ell},+} - \ket{\bar{\ell},-} \bra{\bar{\ell},-} \, \right]
\end{equation}
It is immediately clear that the action of the Hamiltonian on the subspaces of the antisymmetrized and symmetrized states is identical except for the exchange of $t' \longrightarrow -t'$.  We discuss the energies of states in both subspaces in parallel. The Hamiltonian is also symmetric under the parity operator, $P \ket{\ell,\pm} = \ket{-\ell, \pm}$; since only the states symmetric under $P$ will be affected by the crossing point, we focus on these even parity states in the following.   We make an ansatz for the unnormalized, (anti-) symmetrized, even parity eigenstates of the Hamiltonian by writing
\begin{equation}
\label{ansatz}
| k, \pm \rangle = \sum_{\ell = -m}^{m} \cos \left( a k | \ell | + \phi^\pm_k \right) | \ell \rangle
\end{equation}
where $\phi^\pm_k $ represents the wavevector-dependent break in the phase of the wavefunction due to the interchain interaction on chains of $N= 2m + 1$ tight binding sites. Boundary conditions require that the amplitude of the wavefunction at sites $\pm (m+1)$ vanishes, which with Eq.~\ref{ansatz} leads to
\begin{equation}
\label{quantizationI}
a k ( m+1) + \phi^\pm_k = \frac{\pi \left( 2 p + 1\right)}{2}, \hspace{0.2cm} p \in {\cal Z}.
\end{equation}
One can show that the states $| k, \pm \rangle $ with arbitrary phase $\phi^\pm_k$ satisfy the time-independent Schr\"{o}dinger equation all sites away from the crossing point since, via direct calculation, one finds
\begin{equation}
\label{schro1}
E^\pm_k \langle \ell \neq 0, \pm | k, \pm \rangle = \langle \ell \neq 0, \pm | H | k, \pm \rangle,
\end{equation}
with
\begin{equation}
\label{energy}
E^\pm_k = - 2 t \cos ( a k ).
\end{equation}
However, to simultaneously satisfy this eigenvalue equation at the crossing point as well as Eq.~\ref{quantizationI}, we must choose the phase angle $\phi^\pm_k$ such that
\begin{equation}
\label{quantizationII}
\sin ( k a) = \mp \frac{t'}{2 t} \tan \left[ k a ( m+ 1 ) \right].
\end{equation}
The solutions of this transcendental equation that determine the allowed wavevectors $k$ (and energies via Eq.~\ref{energy}) are shown graphically as the set of intersections in figure \ref{quant-fig} where the left and right
hand sides of Eq.~\ref{quantizationII} are plotted.

The principal consequence of the crossing point is the
creation of two localized bound states out of the
symmetrized/anitsymmetrized states $\ket{ k, \pm}$. We examine the
symmetrized states first. The conditions on the tunnelling matrix
element $t'$ for the appearance of this bound state can be
inferred from Eq.~\ref{quantizationII}: If
\begin{equation}
\label{condition}
1 < \frac{|t'|}{2 t} ( m+1)
\end{equation}
then there is no solution for real wavevector $k a  < \pi/[2 (m+1)]$, but a new solution appears along the imaginary axis in the complex $k$ plane at $k = i \kappa$, where $\kappa$ is given by
\begin{equation}
\label{kappa}
\sinh(a \kappa) = \frac{|t'|}{2 t} \tanh\left[a \kappa (m+1)\right].
\end{equation}
From this relationship and Eq.~\ref{energy}, we determine the energy of the resulting bound state to be
\begin{equation}
\label{bound-state}
E_b = - 2 t \sqrt{ 1 + \frac{t'^2}{4 t^2} \tanh^2[\kappa (m+1)] }
\end{equation}
where $\kappa$ is the solution from Eq.~\ref{kappa}.
From the magnitude of the imaginary wavevector, it is clear that this bound state is localized on the length scale of $a t/t' \sim a $. It is important to note that, due to the appearance of this bound state on any chain of reasonable length, the energy of the electronic degrees of freedom is decreased by a quantity on the order of the tunnelling matrix element $t'$. In the limit of an infinite chain where $m \longrightarrow \infty$, one sees from the above equations that $a \kappa \sim {\cal O}(1)$ and the $\tanh(\cdot)$ in Eq.~\ref{bound-state} becomes one in agreement with previous work \cite{economou:79}. This rearrangement of the electronic degrees of freedom has dramatic consequences for the polymer conformational dynamics as we show below.

In addition to this lower energy bound state, a second localized state is pulled from the conduction band of antisymmetrized states. In this case $t' \longrightarrow -t'$ as discussed below Eq.~\ref{diag-interchain}. From Eqs.~\ref{condition},\ref{kappa} we now find that the complex wavevector associated with the bound state becomes $k = i \kappa + \pi$; the bound state appears at the edge of the Brillouin zone and the energy of the state is $- E_b$. It is interesting to note that the appearance of these two localized states at energies $\pm E_b$ is directly analogous to the appearance of bonding/anitbonding states in covalently bonded atoms\cite{Pauling:55}. In the present case, however, we build these states out of spatially extended conduction-band states rather than the localized atomic orbitals.

There are still other consequences of interchain tunnelling for the scattering states that remain in the conduction band. In other words, all the even symmetry, extended states of the unperturbed chains are shifted in energy due to their interaction with the crossing point at site zero. These scattering states remain extended, {\em i.e.} retain a purely real wavevector, but that wavevector shifts due to the interaction with the crossing point. For $t' = 0$ we trivially find these extended states at $k^{(0)}_p = \pi (2 p + 1)/(2 N)$ for integer $p$. Due to scattering at the crossing point, these wavevectors shift so that $k_p \longrightarrow k_p + \Delta k_p$. These shifts can be computed by expanding both sides of Eq.~\ref{quantizationII} near $k^{(0)}_p$. From this expansion we find to leading order in $t'/N$ that
\begin{equation}
\label{shifts}
\Delta k_p = \frac{-t'}{2t N \sin k_p} \left ( 1 + \frac{t'}{2t N} \frac{\cos k_p}{\sin^2 k_p } \right) + {\cal O} \left( \frac{1}{N^3} \right).
\end{equation}
Using the above shifts in $k_p$, we can compute using Eq.~\ref{energy} the sum of the shifts in the energy levels of these scattering states in the symmetrized band. Recalling that we need to also include the analogous shifts in the chain antisymmetrized band, which eliminates terms odd in $t'$, we find that the total shift is
\begin{equation}
\label{e-shifts}
\Delta E = - \left(\frac{t'}{N}\right)^2 \, \sum_{p=1}^{(m_{\rm f})} \frac{\cos k_p}{2t \sin^2 k_p} ,
\end{equation}
where the sum is over all filled energy levels from the bottom of
the conduction band to $m_{\rm f}$ set by the Fermi level of the
system.  Since the sum of all such energy
shifts of the extended states vanishes in the limit of large $N$
as $1/N$, the changes in the energy of the
scattering states in the band is not significant for long chains.  Hereafter we will ignore the finite length ($O(1/N)$) The principal result is that there is an attractive
interchain interaction due to a reorganization of the electronic
degrees of freedom of the system. The energetically dominant part
of this reorganization is the appearance of the ``bonding'' bound
state that lowers the electronic energy essentially by $t'^2/t$.

\begin{figure}[htpb]
  \centering
  \includegraphics[width=8.0cm]{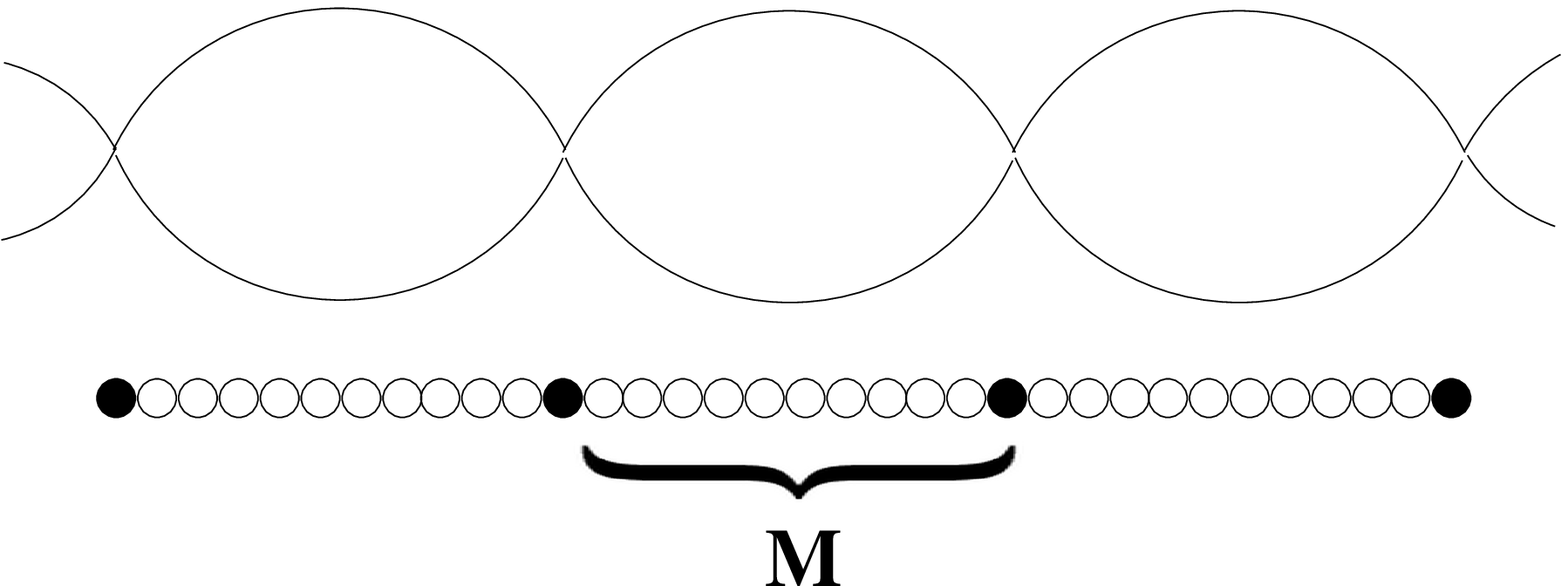}
  \includegraphics[width=8.0cm]{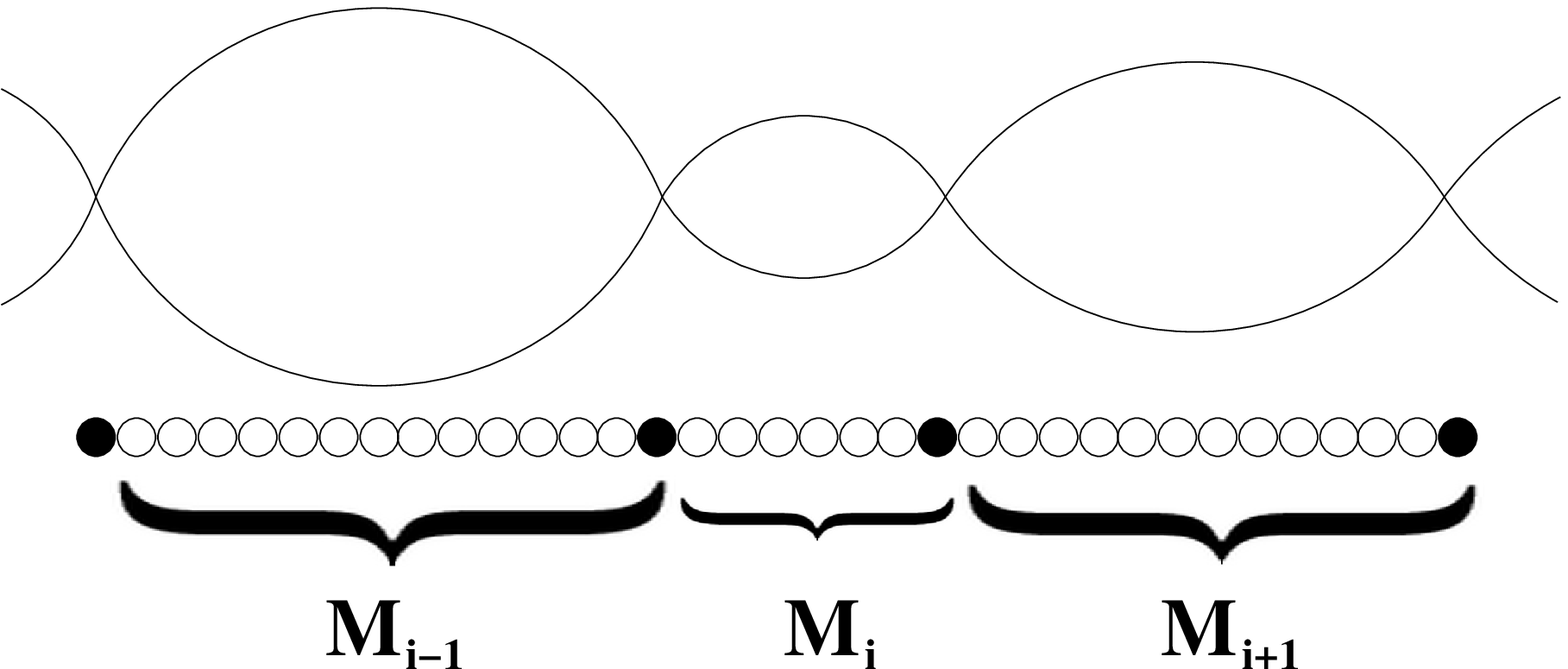}
  \caption{Two polymers interacting at a series of localized sites. In the upper figure the two chains form an ordered structure in which a crossing point appears after exactly $M$ tight binding sites. The lower figure represents the disordered version of the braid. Below each braid is a pictorial representation of the appropriate tight-binding Hamiltonian in which the filled circles represent the interchain tunnelling sites. In the disordered two-chain braid, the number of sites between crossing points varies along the chains, $M \longrightarrow M_i$, but the same $M_i$ sites exist on both chains between the crossing points. }
\label{braidmodel}
\end{figure}

Since the dominant contribution to the electronic energy shifts comes from a localized state at the crossing point,  it is perhaps not surprising, that, when the above analysis is generalized to the case of a crossing point at an arbitrary point along the chain, the principal, qualitative result is unchanged.  We will show below that this basic result is effectively insensitive to interactions between such crossing-points in chains that form a multicrossing-point ``braids''.  We do this in two steps by first considering an ordered array of crossing points  and then by examining a spatially disordered set of crossing points-see figure~\ref{braidmodel}.

\subsection{An Ordered Array of Crossing Points}
\label{ordered}

One may wonder about the effect of a multiple set of interchain crossings. Specifically, if, as it appears above, one crossing point decreases the overall energy of the electronic states of the system, is it more profitable for the chains to form many such crossing points, and if so, at what density?  To address this question, we first consider the most simple such structure -- an ordered array of crossing points at every $M^{\rm th}$ site along the chains.  The ordered structure forms, in effect, a superlattice in which the unit cell is constructed from one crossing point and a basis of $M-1$ tight binding sites between that consecutive crossing points. Qualitatively, one expects the localized states centered at each crossing point to merge into what may be termed an impurity band in the lattice. We explore this point below paying particular attention to the interchain binding energy per unit length of the polymers.

To begin we note that within a single supercell of the superlattice, the tight-binding Hamiltonian becomes
\begin{eqnarray}
\label{supercell-H}
H_n &=& -t \sum_{\ell = 1}^{M-1} \left\{ \, \ket{\ell,n} \bra{\ell+1,n} + \ket{\ell+1,n} \bra{\ell,n} \right\} \nonumber \\
 & & + \lambda \ket{M,n} \bra{M,n}
\end{eqnarray}
where state $\ket{\ell,n}$ is a position eigenstate representing
an electron at the $\ell^{\rm th}$ site in  the $n^{\rm th}$
supercell. Due to the interchain tunnelling at the $M^{\rm th}$
site in each supercell, there is a localized potential for the
electron to reside there. In the above equation and in what
follows we find it convenient to suppress the $\pm$ notation
describing the interchain symmetrized and antisymmetrized states.
We have seen above in Eq.~\ref{diag-interchain} that the only
difference in the form of the Hamiltonian acting on these two
subspaces of opposite chain-exchange symmetry is the shift: $t'
\longrightarrow -t'$. In the above equation, that $t'$ dependent
term in the Hamiltonian is given in terms of $\lambda$ which takes
on the value $t' \, (-t')$ for the antisymmetrized (symmetrized)
states.   To complete our superlattice Hamiltonian, we include a
simple tunnelling term to couple the states in each supercell.
Here, as before, we assume that the presence of the crossing chain
does not affect the intrachain hopping matrix element, although
such effects could be taken into account using the present
tight-binding formalism.  The coupling term is
\begin{equation}
\label{supercell-coupling-H}
H'_n = -t \left\{ \ket{M,n} \bra{1, n+1} + \ket{1,n+1} \bra{M,n} \right\}.
\end{equation}
The first term in the above equation couples the crossing-point
site in the $n^{\rm th}$ supercell to the tight-binding site to
its right (in the $(n+1)^{\rm th}$ supercell) with the usual
intrachain hopping matrix element, while the second term allows
for electron hopping from the first site in $(n+1)^{\rm th}$
supercell onto the crossing point site to its immediate left that
is part of the $n^{\rm th}$ supercell.  The full Hamiltonian of
the system is simply the sum of these two terms summed over all
$N$ supercells:
\begin{equation}
\label{superlattice-full-H}
H= \sum_{n=1}^N \left( H_n +  H'_n \right),
\end{equation}
where we have used periodic boundary conditions so that the $(N+1)^{\rm th}$ supercell is identified with the first supercell.  Since principal effect of the interchain tunnelling on the energetics of the system has been shown to be the appearance of  localized bound states for a single crossing point, the net effect of periodic boundary conditions used here is minimal.

In order to diagonalize the Hamiltonian given by
Eq.~\ref{superlattice-full-H}, we introduce two wavevectors
conjugate to the real-space degrees of freedom that index the
supercell and the site within each supercell by writing
unnormalized states
\begin{equation}
\label{Bloch}
\ket{q,k} = \sum_{n=1}^N \sum_{\ell = 1}^M \exp^{i q a M n} \left( e^{i a k \ell} + A_{q,k} e^{- i k a \ell} \right)  \ket{\ell,n},
\end{equation}
where $A_{q,k}$ will be determined below.  The allowed values of $q$ are fixed by the periodic boundary conditions to be $ q = 2 \pi n/NMa$ where $n$ is an integer such that: $- N/2 < n < N/2$. We now insist that these states, which are eigenstates of the supercell translation operator are also energy eigenstates: $H \ket{q,k} = E(q,k) \ket{q,k}$. This relation can be translated into three related equations formed by projecting the above relation onto three carefully chosen states. We require that
\begin{eqnarray}
\label{req-one}
\bra{\bar{\ell}, \bar{n}} H \ket{q,k} &=& E(q,k)  \langle \bar{\ell}, \bar{n} |q,k \rangle \\
\label{req-two}
\bra{1, \bar{n}} H \ket{q,k} &=& E(q,k)  \langle \bar{\ell}, \bar{n} |q,k \rangle \\
\label{req-three}
\bra{M, \bar{n}} H \ket{q,k} &=& E(q,k)  \langle \bar{\ell}, \bar{n} |q,k \rangle,
\end{eqnarray}
where $\bar{\ell} \neq 1,M$ and the index of the supercell $\bar{n}$ is arbitrary.
The first condition determines the energy values to be $E(q,k) = -2 t \cos (k a)$. Note that since condition Eq. \ref{req-one} applies to the interior of each supercell, the energy eigenvalues appear to be independent of $q$ and $\lambda$. This apparent lack of $q$- and $\lambda$-dependence is false, as we will soon be forced to require that the allowed values of $k$ depend on these parameters. The second and third conditions above acknowledge the special role of the crossing-point in the Hamiltonian. From the second condition, we determine the parameter $A_{q,k}$. In order to satisfy the eigenvector/eigenvalue equation at the left-most site in a supercell, we find
\begin{equation}
\label{A-determined}
A_{q,k} = \frac{e^{i (k a M - q)}- 1}{1 - e^{-i (k a M + q)}}.
\end{equation}
In the above result we have used the energy eigenvalues obtained from enforcing the first condition.
Finally, the combination of the third condition, Eq.~\ref{req-three} along with the energy eigenvalue used above leads to a relation between, $k$, $q$, and $\lambda$ of the form
\begin{equation}
\label{k-q-condition}
\frac{\lambda}{2 t} \tan \left( k M a \right) = - \left[ 1 - \frac{\cos (q M a)}{cos( k M a)} \right] \sin (k a).
\end{equation}
The above equation may be read as an implicit form of the relation of $k$ upon $q$ and $\lambda$, {\em i.e.} $ k = k (q,\lambda)$. Note, that in the limit of a very sparse array of crossing-points, {\em i.e.}  $M \longrightarrow \infty$ we recover the condition on imaginary $k$ for the isolated bound state given in Eq.~\ref{kappa}. However, as $M$ becomes smaller, we find a small modification in this condition and the introduction of a lattice momentum ($q$) dependence in $k$. We now turn to the calculation of the density of states $g(E) = \partial {\cal N}/\partial E$ where ${\cal N}$ is the total number of energy eigenstates with energy less than $E$.  Using the density we compute the total energy of the electronic degrees of freedom of the braided polymers.  The density of states can be written as
\begin{equation}
\label{density-states}
g(E) = \frac{\partial {\cal N} }{\partial k} \cdot \left( \frac{1}{\frac{\partial E}{\partial k}} \right).
\end{equation}
The second derivative in the product above is computed trivially from the dispersion relation. Using the quantization of the wavevector $q$, the first derivative can be expressed as:
\begin{equation}
\label{first-der}
\frac{\partial {\cal N} }{\partial k} = \frac{\partial {\cal N}}{\partial n} \cdot \frac{\partial n}{\partial q} \cdot \frac{\partial q}{\partial k} = M \cdot \frac{N M a}{2 \pi} \cdot \frac{\partial q}{\partial k}
\end{equation}
where the first term in the product reflects that fact that for each wavevector $q = q_n = 2 \pi n/(N a) $ there are $M$ allowed values of $k$, one for each of the $M$ bands in the system. The second term in the above product expresses the $q$-quantization condition; this leaves only the last term to be computed from Eq.~\ref{k-q-condition}.  This derivative is given by
\begin{equation}
\label{dqdk}
\frac{\partial q}{\partial k} = \frac{ \frac{\lambda}{2 t} \left[ -M \cos ( k a M) + \sin(k a M) \frac{\cos(k a)}{ \sin^{2}} (k a) \right]
+ M \sin( k a M)}{M \sqrt{ 1 - \left( \frac{\lambda}{2 t} \sin (k a M) \csc ( k a ) + \cos ( k a M ) \right)^2 } }.
\end{equation}
Now, the total energy of the electronic states of the chains can be computed from Eqs.~\ref{density-states}, \ref{first-der}, and \ref{dqdk}. 

\begin{figure}[htpb]
  \centering
  \includegraphics[width=8.0cm]{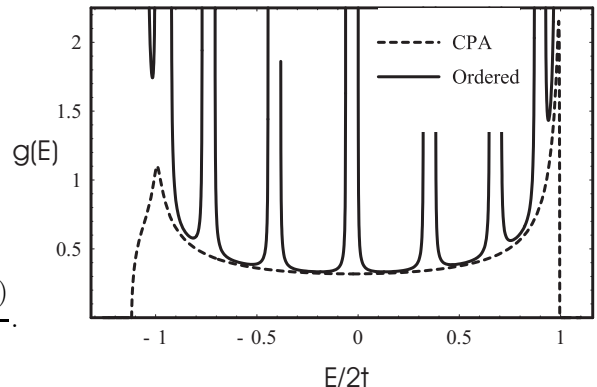}
  \caption{The density of states as a function of energy for an ordered array of crossing points (solid line) and a disordered array of crossing points (dashed line). In both cases the density of binding sites is $p = 1/M = 0.125$ and $\lambda/(2 t) = -0.25$.}
\label{density-states-plot}
\end{figure}

One expects from the calculation of the isolated crossing-point that the change in electronic energy is almost entirely due to the creation of a narrow band of bonding states below the conduction band so it is reasonable to imagine that the total change in electronic energy is independent of the Fermi level. For convenience we choose the Fermi level to be $E_{\rm F} = 0$.   Since there is at most one energy eigenstate for each (complex) value of $k$, we may write the total energy of the electronic system as an integral over $k$ of the density of states and the dispersion relation.  All possible states below the Fermi level are explored by integrating along the following contour in the complex plane: ${\cal C} = (-i \infty, 0) \bigcup (0, \pi/2)$ which is the combination of the negative imaginary axis and the segment of the real axis spanning 0 to $\pi/2$.  The density of states along this contour is shown in figure \ref{density-states-plot} as a function of energy. Lastly, using Eq.~\ref{k-q-condition} we note that for there to be an energy eigenstate with a given $(q,k)$, it must be true that $| \lambda/(2t) \sin(M a k) \csc( a k ) + \cos (k a M) | < 1$ so that the denominator of Eq.~\ref{dqdk} must be real. To enforce this condition over the $k$ integration we take the real part of the integral to write
\begin{equation}
\label{energy-final}
E = \mbox{ Re } \int_{\cal C} \frac{\partial {\cal N}}{\partial k} E(k) dk .
\end{equation}

The change in the total electronic energy as a function of $t'$
and $M$ can then be computed; these results are shown in figures
\ref{threemodelst} and \ref{binding-e-plot}.  The observed even-odd effect in figure \ref{threemodelst} is
an artifact of our choice of the Fermi level.  For chains with
even $M$, a band gap will open at $E_F=0$ resulting in an
additional reduction in the energy.  In practice, we expect that
the Fermi level of doped polymers will not lie at a place of such high
symmetry and the dominant contribution to the energy will be the
creation of localized bound states.

The principal result
of this calculation is that, due to the highly localized nature of
the bonding states at each crossing-point, the total interchain
binding energy is essentially a linear function of the crossing
point density. In other words the attractive interaction at each
crossing point is highly independent of the local density of such
crossing points at least until $M$ decreases to order unity.  

\subsection{A Disordered Array of Crossing Points}
\label{disordered}

As a final point regarding the electronic contribution to the interchain adhesion, we now allow the set of crossing points to become disordered in a specific manner as shown in the lower part of figure \ref{braidmodel}. We allow the number of tight-binding sites between two crossing points ($M$) now to vary along the chain in an uncorrelated manner so that the chains may be described by a sequences of such spacings $M$, $M_1, M_2, \ldots, M_p$. We restrict the form of the randomness to cases in which the {\em same} sequence can be used to describe both chains. The advantage of this restriction is that the interchain interaction part of the Hamiltonian is still diagonal in the basis of chain symmetrized/antisymmetrized wavefunctions.

When considering  the chain-symmetrized/antisymmetrized subspaces individually, the Hamiltonian given by Eqs.~\ref{tb-interchain},\ref{tb-interaction} is identical to the two-component random lattice problem, which has been extensively studied \cite{Schmidt:57,Soven:67,Velicky:68,Soven:69}.  For completeness we recapitulate the discussion of the general formalism of the coherent potential approximation (CPA) developed therein while applying that formalism to the system at hand. We begin by expanding the propagator of the full, disordered system $G$ in terms of the propagator of a uniform system $G_0$ with a fixed, and as yet unspecified on-site potential.  The propagator for the uniform system is given by
\begin{equation}
\label{propagator-uniform}
G_0 = \frac{1}{E - H_0}
\end{equation}
where
\begin{equation}
\label{uniform-H}
H_0 = -t \sum_{\ell=1}^N \left( |\ell + 1\rangle \langle \ell | + |\ell \rangle \langle \ell +1 | \right) + v \sum_{\ell =1 }^N \ket{\ell} \bra{\ell}
\end{equation}
and $v$ is the spatially independent on-site energy. The full Hamiltonian of the disordered system is given by the sum of $H_0$ from Eq.~\ref{uniform-H} and scattering potentials at each site given by
\begin{equation}
\label{random-int}
H_I = \sum_{\ell =1 }^N (\varepsilon_\ell - v)  \ket{\ell} \bra{\ell}.
\end{equation}
In the above equation, the random variable $\varepsilon_\ell$  introduces the quenched disorder by introducing uncorrelated tunnelling sites with density $p$ along the chain.  This random variable is takes the value $\lambda$ with probability $p$ and is zero otherwise where, as before, $\lambda$ is $\pm t'$ for the chain-antisymmetrized/symmetrized Hamiltonian.

One can show in the usual way that the propagator for the full, random system $G$ can be expressed perturbatively in terms of the tunnelling Hamiltonian $H_I$ and propagator of the uniform system so that
\begin{equation}
\label{full-propagator}
G = \frac{1}{E - H} = G_0 + G_0 H_I G_0 + \cdots \, .
\end{equation}
Our goal, as before is to compute the density of states of the system, g(E), which is given in terms of the propagator as
\begin{equation}
\label{DOS}
g(E)  = - \frac{1}{\pi} \mbox{Im} \sum_\ell \bra{\ell} G \ket{\ell}
\end{equation}
where the sum is over a complete set of states. To approximate this sum accurately, we follow the work of Soven \cite{Soven:69} and first reorganize the perturbation series given in Eq.~\ref{full-propagator} by introducing the $T$ operator representing the Born-series for scattering off a particular site, $\ell$,
\begin{eqnarray}
\label{T-operator}
T_\ell  &=& \ket{\ell} (\varepsilon_\ell - v) \bra{\ell} + (\varepsilon_\ell - v)^2 \ket{\ell} \bra{\ell} G_0 \ket{\ell} \bra{\ell}  + \cdots \\ \nonumber
& =& \frac{1}{1 - \bra{\ell} G_0 \ket{\ell} (\varepsilon_\ell - v)} \ket{\ell} \bra{\ell},
\end{eqnarray}
so that Eq.~\ref{full-propagator} implies
\begin{eqnarray}
\label{reorganized}
\bra{\ell'} G \ket{\ell} &=& \bra{\ell'} G_0 \ket{\ell} + \sum_{\bar{\ell}} \bra{\ell'} G_0 \ket{\bar{\ell}}T_{\bar{\ell}} \bra{\bar{\ell}} G_0 \ket{\ell} + \\ \nonumber
+&  & \sum_{\bar{\ell} \neq \bar{\bar{\ell}}}  \bra{\ell'} G_0 \ket{\bar{\ell}}T_{\bar{\ell}} \bra{\bar{\ell}} G_0 \ket{\bar{\bar{\ell}}} T_{\bar{\bar{\ell}}}\bra{\bar{\bar{\ell}}} G_0 \ket{\ell} + \cdots \, .
\end{eqnarray}
The successive terms in the above equation represent the propagator of the electron in the uniform system followed by a correction due to the interaction of that electron with the scattering potential at site $\bar{\ell}$, followed in turn by the interaction of that electron with sites $\bar{\ell}$ and $\bar{\bar{\ell}}$; the higher order terms (not shown above) have an analogous interpretation. It is important to note that, due to the reorganization the perturbation series in terms of the $T$ operators, the effect of the electron interacting with the {\em same} site successively has been already taken into account. Thus, in the sums representing the second and higher order terms it is necessary to restrict the summation to avoid revisiting the same site twice in a row, {\em e.g.} $\bar{\ell} \neq \bar{\bar{\ell}}$ in the second order term above.

\begin{figure}
\centering
  \includegraphics[width=8.0cm]{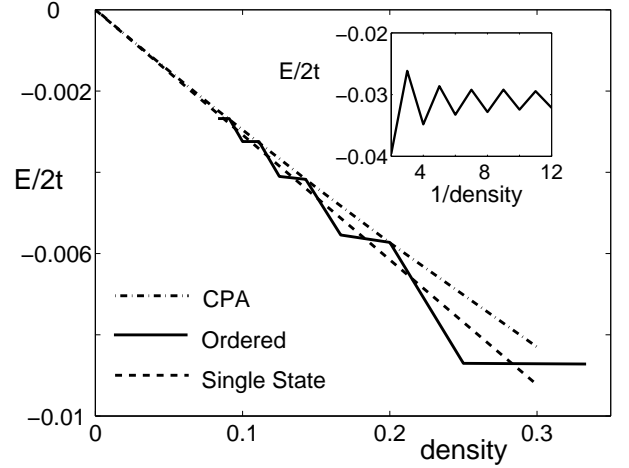}
\caption{Plot of the change in electron energy as a function of linker density
for cases of quenched, disordered positions (short dashes) and ordered positions (solid line).
By assuming the independence of the binding sites so that each linker contributes $4 t^2-(4 t^2 +t'^2)^{1/2}$, one
arrives at a reasonable linear approximation to the binding energy as a function of cross-link density (long dashes).  Inset:  Plot of the change in energy for an ordered lattice vs. 1/density.
Both plots use $\lambda/2t =0.25$.}
\label{threemodelst}
\end{figure}

We now choose the heretofore arbitrary on-site potential. One
might imagine that it would be reasonable to choose that potential
to simply be the mean of the on-site energies of the disordered
system by taking $v = p \lambda $, however, we are not attempting
to average the Hamiltonian over the disorder but rather we will be
averaging the propagator. Thus, using the CPA approach, we choose
that on-site potential to enhance the convergence of the
perturbation series for $G$ given in Eq.~\ref{reorganized} by
taking $v$ such that the average of the $T$ operator over the
quenched disorder vanishes.  Representing the disorder averages by
$[\cdot ]$, we choose $v$ such that
\begin{equation}
\label{v-picker}
\left[ \bra \ell T_{\ell} \ket{\ell} \right] = (1-p) \frac{-v }{1 + v \bra{\ell} G_0 \ket{\ell} } + p \frac{\lambda - v}{1 - (\lambda - v)  \bra{\ell} G_0 \ket{\ell}} = 0.
\end{equation}
Now, the reorganized perturbation series in Eq.~\ref{reorganized}
is greatly simplified. Since $\left[ \bra \ell T_{\ell} \ket{\ell}
\right] = 0$ and  $T$ operators at different sites are
uncorrelated, $\left[ \bra \ell T_{\ell} \ket{\ell} \bra{\ell'}
T_{\ell'} \ket{\ell'} \right] = \left[ \bra{\ell} T_{\ell}
\ket{\ell} \right] \cdot \left[\bra{\ell'} T_{\ell'} \ket{\ell'}
\right] = 0  $, we see that first correction to $G_0$ is fourth
order in the $T$ operators.  In the limit of weak scattering,
$\left[G \right] \simeq G_0$ (with the advantageous choice for $v$
discussed above) makes an excellent approximation since the first
correction is of order $(p \lambda/(2t))^4$.  

Finding the density of states in the disordered system requires
that we evaluate the trace of the disorder--averaged propagator
(Eq.~\ref{DOS}) as a function of $v$; we do this in the position
representation by writing propagator in the uniform system as
\begin{equation}
\label{prop}
\bra{\ell } G_0\ket{m} = G_0(\ell, m;E ) = \frac{a}{2 \pi} \int_{\pi/a}^{\pi/a} dk \frac{e^{ik a (\ell - n) }}{E - v + 2 t \cos (k a) }
\end{equation}
and extracting the diagonal element
\begin{equation}
\label{diagonal}
G_0(\ell, \ell;E ) = \pm \left\{ (E-v)^2 - 4 t^2 \right)^{-1/2}
\end{equation}
and the sign above is determined by whether the magnitude of
$(E-v)/2t$ is greater or smaller than unity. In the uniform system the sum on
position eigenstates is trivial as seen by the $\ell$-independence
of the RHS of the above equation.  However, determining $G_0(\ell,
\ell;E ) $ and $v$ simultaneously requires that we solve two
polynomial equations given by Eqs.~\ref{v-picker} and
\ref{diagonal}.

\begin{figure}
\centering
  \includegraphics[width=8.0cm]{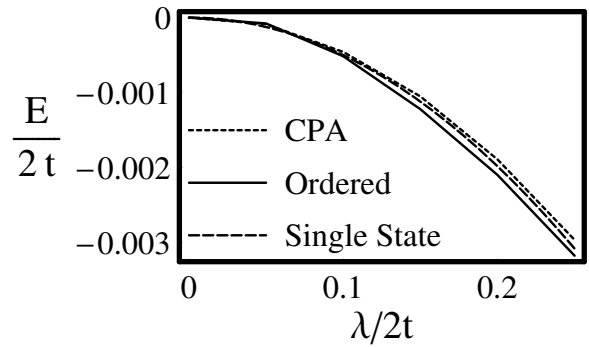}
\caption{Plot of the change in electron energy as a function of the interaction strength
for random disorder (short dashes), perfectly ordered potential (solid line),
and under the independent-linker assumption where each cross-link contributes $4t^2-(4t^2+t'^2)^{1/2}$ (long dashes).  Here the cross-link density is $0.1$.}
\label{binding-e-plot}
\end{figure}

The binding energy is determined by integrating Eq.~\ref{DOS} over the filled states.  For convenience, we chose the Fermi level to again lie at $E_F=0$.  The results are shown in Figs. \ref{threemodelst}, \ref{binding-e-plot}.  As with the ordered lattice of binding sites, we find that the binding energy of the disordered braided chains is very nearly a linear function of the density of sites.

We can conclude that the primary effect of introducing binding points between the polymers is the creation of ``impurity'' bands centered at the energy given by Eq. \ref{bound-state}.  The states in the impurity bands are created at the expense of states near the band edges, $E=\pm 2t$.  However, the density of states far from the band edges is essentially unchanged.  This can be easily proved for a random arrangement of the impurities.  We first note that the density of states is given by
\begin{eqnarray}
g(E)&=&\frac{1}{\pi}\frac{\partial \Phi}{\partial E}  \\
&=&\frac{1}{\pi}\frac{\partial k}{\partial E}\left(N+Np\frac{\partial \overline{\theta}}{\partial k}\right),
\end{eqnarray}
where $\Phi$ is the phase of the wavefunction at the last site on the chain, $p$ is the density of impurities, and $\overline{\theta}$ is the average phase-shift as the wavefunction passes through an impurity.  For an impurity $\lambda$ at the origin $(\ell =0)$, the wavefunction takes the form
\begin{equation}
\ket{k}=\sum_{l\leq 0} \cos (kla +\phi)\ket{l}+A\sum_{l> 0} \cos
(kla +\phi+\theta)\ket{l},
\end{equation}
where $A$ is an amplitude to be determined.  The phase-shift is given by
\begin{equation}
\theta=\tan^{-1}\left(\tan(\phi)-\frac{\lambda}{t\sin(ka)}\right)-\phi.
\end{equation}
Averaging over the incident phase, $\phi$, we find the average phase-shift
\begin{equation}
\overline{\theta}=\frac{t \lambda \cos(ka)}{4t^2 \sin^2 (ka)+
\lambda^2}.
\end{equation}
This shift is odd in the impurity strength, $\lambda$, therefore
the change in the density of states from the symmetric and
anti-symmetric bands will exactly cancel.  Therefore, the
reorganization of electronic energy must occur where this argument
breaks down, namely when the separation between binding sites is
comparable to the wavelength, {\em i.e.} when $\sin (ka)\simeq 0$.  At
this region near the band edges states are removed to form the
imaginary wavenumber impurity states.

The role of the impurity band as the source of the binding energy
is further supported by the ``single state" or independent binding site approximation (long dashes) line in
figures \ref{threemodelst} and \ref{binding-e-plot}.  This line is
the binding energy that would be obtained if the sole effect of
adding a binding site were to move a single electron from the
bottom of the band $(E=-2t)$ to the energy of the isolated bound
state $(E_b=-(4t^2+t'^2)^{1/2})$.  This line is an excellent
approximation for the binding energy of both the ordered lattice
of binding sites (solid line) and the random array of binding
sites (short dashes) which supports the suggestion that the
creation of the impurity band is the dominant effect in the
reorganization of electronic energy.

\subsection{Parallel Configurations}
\label{railroad}

If the chains adopt a parallel configuration, then there will be an overlap between
each site and the corresponding site on the opposite chain.  This interaction is
easily diagonalized with the transformation given by Eq.~\ref{symm}.  This yields the dispersion
relation
\begin{equation}
E_{\pm}(k)=-2t\cos{k}\pm t'.
\end{equation}

We can solve for the binding energy of the polymers in this
zipped configuration if the polymers are lightly doped (a
general solution for the zipped configuration eigenstates may be
found in Appendix A). If each polymer has $\delta/2$ electrons and
$\delta \ll N$ then, prior to zipping, all electrons will have
energy very near $-2t$. If the polymers bind over a length of $m$
monomers, then states are created with energy $E<-2t$ with a
density of states $m(4t^2-(E+t')^2)^{-1/2}/\pi$. The electrons in
these states have lowered their energy by
\begin{eqnarray}
\Delta E&=&-\frac{m}{\pi}\sqrt{4t^2-(-2t+t')^2} \nonumber \\
&&-\frac{m}{\pi}(2t-t')\left(\sin^{-1}(\frac{t'}{2t}-1)+\frac{\pi}{2}\right).
\label{zippedE}
\end{eqnarray}
This corresponds to a binding energy of $\simeq-2t'^{3/2}/3\pi \sqrt{t}$ per site.

If the chains zip past a critical length $m_c=\delta\pi/\cos^{-1}(t'/2t-1)$
then all the electrons have energy $<-2t$.  These electrons now have imaginary
wavenumbers in the unzipped region of the chain, therefore the unzipped regions will
no longer be metallic.  The binding energy is now
\begin{equation}
\Delta E=-\frac{2tm}{\pi}\sin (\frac{\delta \pi}{m})-\delta t'+2t\delta.
\label{zipenergy}
\end{equation}
Since the derivative of this binding energy with respect to the length of the binding region ($m$) is negative, we note that the electrons exert an effective pressure to increase the size of the
zipped region \cite{Pincus:87}. When they form, we expect the length of the zipped regions to be limited only by the number of conduction electrons.   

The binding energy per binding site in a zipped region is $\sim t'^{3/2} t^{-1/2}$ whereas in the braided configurations the binding sites reduce the energy by $\sim t'^2 t^{-1}$. The zipped configuration exhibits stronger binding per site by $\sqrt{t/t'}$; since generically $t' < t$, the electronic degrees of freedom favor zipped configurations of the chains over braided ones. To determine whether a pair of chains will, however, form such zipped regions or the random braids, we must consider the energetics associated with the conformational degrees of freedom of these charged polymers in water.

\section{Conformational degrees of freedom of the chains and equilibrium structures}
\label{polymer}

The equilibrium state of aggregating polymers depends on
the combination of the binding energy due to the electronic degrees of freedom
and the change in the translational/conformation free energy of
the polymer.  From the interplay of these two contributions to the total free energy, we explore the transition from free chains in solution to chain aggregates and discuss their equilibrium structure.  We compare the free energies of (i) isolated chains in solution, (ii) braided pairs of chains consisting of unbound loops between isolated crossing points, and (iii) parallel or ``zipped" configurations in which the chains have numerous consecutive tunnelling sites along their length.  The aggregation of DNA in the presence of divalent ions has been approached in a similar manner\cite{Borukhov:01}, however, in the present system there is no quantity analogous to the fixed concentration of linking divalent ions since the interchain binding sites are created spontaneously at chain intersections.

We begin by considering the coexistence of isolated chains and paired ones in solution.  The thermodynamic potential of a solution with a concentration of chains $n$ can be written up to a trivial constant as 
\begin{equation}
\frac{F(n_2)}{k_BT}=(n-2n_2)(\ln{(n-2n_2)}-1)+n_2(\ln{n_2}-1)+n_2\frac{{\cal E}_b}{k_BT}+n_2 \frac{S_0}{k_B},
\label{dimerenergy}
\end{equation}
in terms of the concentration of bound-chain pairs $n_2$\cite{Landau:80}.  In the above expression $S_0$ is the chain configurational entropy of one polymer: $S_0 \propto k_B L/l_p$.  Its appearance in Eq. \ref{dimerenergy} comes from the fact that in order to pair two chains, one of them must adopt the same configuration as the other and thereby lose its conformational entropy.  The first two terms of Eq. \ref{dimerenergy} represent only the translational entropy of the isolated chains and bound pairs; they do not incorporate this effect.  Minimizing Eq. \ref{dimerenergy}, we find the concentration of dimers is $n^2 \exp ((-TS_0-{\cal E}_b)/k_BT)$.  It remains now to calculate the interchain binding energy ${\cal E}_b$, which is a combination of electrostatic chain repulsion and chain binding due to the proposed tunnelling mechanism.  We do not include van der Waals interactions between chains. This longer-ranged attractive interaction
is generically stronger for the conducting polymers since the
extended electronic states enhance their low-frequency
polarizability.  At higher frequencies one expects
that the polarizability at higher frequencies is typical of small
organic molecules since the dielectric spectrum at higher
frequencies is primarily due to localized molecular states that
are similar to all hydrocarbons. Such van der Waals interactions
lead to a longer range attraction that, in high salt concentrations at
least, are a subdominant correction to the interchain binding
potential when the interchain distance is on the order of
Angstroms. In addition the enhancement of the Van der Waals
interaction is significant for only the metallic rather than the
semiconducting state of the polymers. Thus we expect that the interchain
tunnelling mechanism discussed in this article to dominate the
binding energy of the observed bundles.  We examine this point further in appendix B.

We first consider the binding energy of the more loosely bound, braided configuration of the molecules, ${\cal E}_b^{braid}$.  The binding energy is a combination of the electrostatic interaction of the chains, the bending of the molecules on scales short compared to their persistence length, and the electronic binding energy of the crossing point.  At high salt the first of these terms is significant only near the crossing points.  There the charged chains can be treated as crossed straight rods.  If two charged rods cross at an angle $\theta$ with a
distance of closest approach $d$ the electrostatic energy is
\begin{equation}
E_{ES}=\frac{2\pi k_B T l_B e^{-\kappa d}}{\kappa a_c^2 \sin{\theta}},
\end{equation}
where $l_B=e^2/\epsilon k_B T$ is the Bjerrum length, $a_c$ is the average separation
between charges along the backbone, and $\kappa^{-1}$ is the Debye screening length.

The bending energy of a polymer is
\begin{equation}
E_{bend}=\frac{k_B T l_p}{2}\int \frac{dl}{R(l)^2},
\end{equation}
where $l_p$ is the persistence length, and $R(l)$ is the local radius of curvature.
If the separation between interaction points is comparable to the persistence length,
then in the limit $\kappa^{-1}\ll l_p$ the crossing angle will be very close to $\pi/2$
and the arc the polymers follow will be very nearly circular.  Then the energy per length of the polymers, $L$, is
\begin{equation}
\label{braid-energy}
E_{braid}(l)/L=(\frac{2\pi k_B T l_B e^{-\kappa d}}{\kappa a_c^2}-\frac{t'^2}{2t})/l+\frac{k_B T l_p\pi^2}{4l^2},
\end{equation}
where $l$ is the arc length of the polymer between intersections.  Minimizing with respect to $l$, the energy becomes
\begin{equation}
{\cal E}_b^{braid}=-\frac{(\frac{2\pi k_B T l_B e^{-\kappa d}}{\kappa a_c^2}-\frac{t'^2}{2t})^2}{k_B T l_p\pi^2}L.
\label{Ebraid}
\end{equation}
Here we have neglected the configurational entropy of the loops of chain making up the braid.  As long as $l<l_p$ we expect this entropic contribution to be negligible.

We now turn to the zipped configuration of the chains.  To consider the electrostatic interaction between the two chains, we note that the potential, $\Phi$, around a thin charged rod in a salt solution is given by
\begin{equation}
\Phi(r)=\frac{2l_B k_B T}{a_c}K_0(\kappa r),
\label{potential}
\end{equation}
so the energy of two polymers separated by a distance $d$ over a length $L$ is
\begin{equation}
{\cal E}_b^{zip}(L)= L \frac{2l_B k_B T}{a_c^2}K_0(\kappa d)-L\frac{2t}{\pi a_t}\sin \left(\frac{\delta \pi a_t}{L}\right)-t'\delta+2t\delta
\label{Ezip}
\end{equation}
where $K_0$ is the modified Bessel function, the distance between tight-binding sites is $a_t$, and we have assumed the binding energy is given by Eq.\ \ref{zipenergy}.

We observe that both Eqs. \ref{Ebraid} and \ref{Ezip} depend on the length over which the binding occurs.  So, before we use either of these expressions in Eq. \ref{dimerenergy} we must choose the suitable length, $L^\star$.  For braided chains $L^\star$ is the length of the entire chain and for zipped chains $L^\star$ is determined from \ref{Ezip} yielding $L^\star\simeq 2t\delta a_c^2/l_B k_B T K_0(\kappa d)$.  Note that $S_0$ in Eq. \ref{dimerenergy} is also proportional to $L^\star$.

\begin{figure}
\centering
\includegraphics[width=8.0cm]{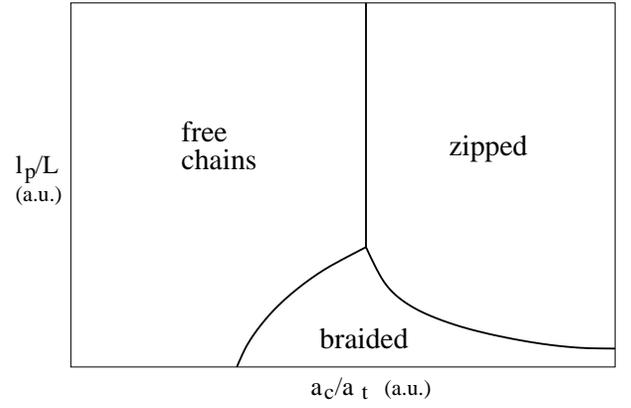}
\caption{Schematic phase diagram for two charged, conducting
polymers. This diagram is spanned on the vertical axis by the
persistence length scaled by the total contour length of the chain
while the horizontal axis is the dimensionless ratio of the the
distance between charges along the chain $a_c$ to the distance
between potential tunnelling sites $a_t$.  In the upper left
region of the figure, chains of high charge density and long
persistence length are unbound in solution (free); upon reducing
the linear charge density along the chain and moving to the right
in the diagram the chains generically become bound in a parallel
or zipped configuration. For flexible enough chains of
intermediate charge density, however, more loose, braided
configurations are predicted.} \label{phase-diagram}
\end{figure}

Based on these considerations we propose a schematic phase diagram for charged, conducting polymers in solution. This diagram (shown in figure \ref{phase-diagram}), which represents the range of phase behavior of these polymers, is spanned by the persistence length and the (suitably scaled) inverse, linear charge density of the chains. This inverse charge density is measured in terms of the ratio of the distance $a_c$ between charged groups along the polymer backbone divided by the distance between potential tunnelling sites, $a_t$. In the upper right portion of the figure the chains are both stiff and have a high charge density relative to their density of potential binding sites. Thus, they are unbound in equilibrium.  Generally, moving down and to the right in the diagram induces intermolecular binding due the mechanism presented in this work. The equilibrium bound configurations take the form of parallel, tightly bound pairs, which we denote as zipped chains or as more loosely bound braids.

The onset of braiding is determined by setting that contour length between binding sites equal to the total contour length of the chain, $L$. This gives the free-chains/braid boundary as
\begin{equation}
\label{free-braid-boundary}
\frac{\ell_p}{L} = \frac{2}{k_{\rm B} T \pi^2} \left( \frac{t'^2}{2 t} - \frac{ 2 \pi k_{\rm B} T \ell_p e^{-\kappa d}}{\kappa a_c^2 } \right).
\end{equation}
The term in parenthesis on the right-hand side of the above
equation is collectively the effective binding of one interchain
crossing. The first part of that term is the binding energy due to
the reorganization of the electronic states of the polymers while
the second term represents the reduction in binding energy
associated with the electrostatic interaction of the two chains.
For the cases of current interest in which a single interchain
bond is favorable the right-hand side of
Eq.~\ref{free-braid-boundary} is positive.

To explore the  boundary between zipped and braided states of the polymers we compare the energy of the braided configuration given by Eq.~\ref{Ebraid} to the energy of the zipped configuration. This latter energy is a combination of the binding energy of a zipped region given by Eq. \ref{Ezip}.  For the case of lightly doped chains so that contour length of the zipped regions $L^\star$ is smaller than that of the whole polymer, one finds that the boundary between braided and zipped chain configurations to occur where
\begin{equation}
\label{braid-zip} \frac{\ell_p}{L} = \frac{\left( \frac{t'^2}{2 t}
- \frac{2 \pi k_{\rm B} T \ell_p e^{-\kappa d}}{\kappa a_c^2}
\right)^2}{ L^\star k_{\rm B} T \pi^2 \left( \frac{2 t'^{3/2}}{2
\pi a_t t^{1/2}} - 2 \ell_p k_{\rm B} T K_0(\kappa d) a_c^{-2}
\right) }.
\end{equation}
This phase boundary continues to smaller $a_c/a_t$ until it intersects the free/braid boundary discussed above. There it terminates. Finally, there is a vertical line separating free and zipped chains that intercepts the intersection point of the free/braid and braid/zipped boundaries. This last separation is independent of chain persistence length since the energy of neither the zipped configurations nor the free chains depends on the persistence length.

\section{Discussion}
\label{conclusions}

We find that conducting polymers in the metallic state, {\em i.e.} where
the Fermi level lies within a band, will generically aggregate in
solution due to the reorganization of the electronic degrees of
freedom of these molecules upon the close (Angstrom-scale)
approach to each other. This reorganization of the extended
electronic states of the molecule results in the formation of
localized bound states at these crossing-points and the consequent
reduction of the electronic energy more that compensates for the
reduction of chain configurational entropy due to the creation of
the cross-links between the two polymers. It is interesting to note
that these localized bound states are the direct analogs of the
standard binding/anti-binding orbitals that are created out of
atomic electronic states upon the close approach of their
respective nuclei. In the case of conducting polymers, however,
these localized bound states are created out of the extended
states of the metallic molecules.

The localized nature of these states has the consequence that the binding energy is very nearly linear in the density of binding sites when these sites are separated by at least a few monomers as they are in all braided configurations. However, when the polymers lie in parallel, zipped configurations, the resulting binding energy per site is stronger than the isolated binding sites by $O((t/t')^{1/2})$.  The equilibrium structures formed by these polymers will depend not only on the interchain binding mechanism, but also the linear charge density of the polymers, solvent screening, and the persistence length of the polymer.  The combination of these factors will determine whether the polymers form loosely braided structures, or tight bundles.

Based on these considerations we have proposed a schematic phase diagram showing the expected equilibrium structures that may be observed in such systems.  For large enough linear charge density, one finds that the lowest energy state of the aggregate is a braid consisting of loops of unbound polymer between cross-linking binding sites. For smaller linear charge density one finds that the energetically favorable configuration is two parallel chains where the interchain distance remains on the order of a few Angstroms all along their arc length.

Clearly, the direct experimental observation of such structures is
difficult. We note that SANS experiments \cite{Wang:01,Wang:01b}
are consistent with the formation of bundles in aqueous solutions, but
the spatial resolution of these experiments is not fine enough to
distinguish between the braid and parallel phases discussed above.
Moreover, the current calculations are not directly applicable to
these experiments since previous experimental studies focussed on
semiconducting polymers. Nevertheless, our proposed binding mechanism mediated by the reorganization of the electronic degrees of freedom is also relevant to the semiconducting case\cite{Schmit:04}.  Since it will likely remain difficult to quantitatively test the detailed structure of the aggregate one must directly probe the electronic states by studying the spectroscopic signature of the molecules as they aggregate.

\acknowledgements
The authors would like to thank F.\ Pincus, G.\ Bazan, and A.J.\ Heeger for stimulating conversations. JDS would also like to thank D.\ Scalapino and L.\ Balents for helpful discussions. JDS acknowledges the hospitality of the University of Massachusetts, Amherst.  This work was supported in part by the MRSEC Program of NSF DMR00-80034.

\appendix

\section{Zipped configuration states}
Here we solve for the eigenstates for two chains that are zipped together over a portion
of their length.  The Hamiltonian is
\begin{equation}
H_0 = -t \sum_{\ell=1}^{N-1} \left( |\ell + 1\rangle \langle \ell | + |\ell \rangle \langle \ell +1 | \right) + \sum_{\ell =1 }^{N-1} v_l\ket{\ell} \bra{\ell}
\end{equation}
where $v_l=0$ in the unzipped regions ($1\leq l \leq n$ and $n+m+1
\leq l \leq N$) and $v_l=\lambda$ in the zipped region ($n+1 \leq l
\leq n+m$).  Here $\lambda=\mp t'$ for the symmetric
(anti-symmetric) states, $m$ is the number of sites in the
zipped region, and $n(n')$ is the number of sites in the unzipped region to the left(right) of the zipped region.  The chains have a total length of $N=n+m+n'$.  The eigenfunctions are of the form
\begin{eqnarray}
\ket{k}=&&\sum_{l=1}^n \sin (lk) \ket{l}+A\sum_{l=n+1}^{n+m} \sin (lk'+\phi)\ket{l} \nonumber \\
&&+B\sum_{l=n+m+1}^{N} \sin (lk+\theta)\ket{l},
\end{eqnarray}
where $A$ and $B$ are undetermined amplitudes, and $-2t\cos k=-2t\cos k' +\lambda$.
The boundary conditions are
\begin{eqnarray}
\sin ((n+1)k)&=&A\sin ((n+1)k'+\phi) \nonumber \\
\sin (nk)&=&A\sin (nk'+\phi) \nonumber \\
A\sin((n+m+1)k'+\phi)&=&B\sin((n+m+1)k+\theta)  \nonumber\\
A\sin((n+m)k'+\phi)&=&B\sin((n+m)k+\theta)  \nonumber \\
(N+1)k+\theta&=&j\pi,
\end{eqnarray}
where $j$ is an integer.  After some algebra, we find two equations for $k$ and $\phi$
\begin{eqnarray}
\frac{\sin (k n')}{\sin(k(n'+1))}&=&\frac{\sin(k'(n+m+1)+\phi)}{\sin(k'(n+m)+\phi)} \\
\cot (k'n+\phi)&=&\frac{\cot(kn)\sin (k)-\lambda}{\sin (k')}.
\end{eqnarray}
From these equations we find the quantization condition for $k$
\begin{equation}
\xi \xi'+(\xi+\xi')\sin(k')\cot (k'(m+1))-\sin^2 (k')=0,
\label{zippedquant}
\end{equation}
where
\begin{eqnarray}
\xi&\equiv&\sin (k) \cot(nk)-\lambda \nonumber \\
\xi'&\equiv&\sin (k) \cot(n'k)-\lambda.
\label{xi}
\end{eqnarray}
From these conditions (Eqs. \ref{zippedquant}, \ref{xi}) and the dispersion relation
\begin{equation}
E_k=-2t\cos (k),
\end{equation}
one obtains the energy eigenvalues.

\section{Intermolecular Van der Waals Attraction}
The van der Waals interaction energy between two like molecules in a screened medium
is given by
\begin{equation}
E_{vdw}\simeq -I\frac{\alpha^2_0 e^{-2 \kappa r}}{\epsilon^2 r^6},
\end{equation}
where $\alpha_0$ is the polarizability of the molecules,
$\epsilon$ is the dielectric constant of the medium, $\kappa^{-1}$
is the Debye screening length, and $I$ is the ionization energy
of the molecules\cite{Israelachvili:92}.  At wide separations the polymers can be
approximated as conducting spheres with a radius $R_g$, thus $\alpha_0 \simeq R_g^3$.  In
aqueous solutions $\epsilon \simeq 80$ and for an $I$ of
order eV then $E_{vdw}\ll 1eV$ since $r\gg R_g$.   If, on the other hand, $r\lesssim R_g$ it is more
accurate to describe the polymers as narrow ellipsoids. In this
situation we approximate the polymers as ellipsoids of length
$l_p$ and thickness $a$ that is a monomeric dimension. The polarizability along the wire is
found to be of order\cite{Landau:84}.
\begin{equation}
\alpha_{\parallel}\simeq \frac{l_p^3}{\ln{\frac{l_p}{a}}},
\end{equation}
and perpendicular to the wire it is
\begin{equation}
\alpha_{\perp}\simeq a^2 l_p.
\end{equation}
So, the interaction energy for parallel and perpendicular wires is
\begin{eqnarray}
E_{\parallel} & \simeq & I \frac{l_p^6 e^{-2\kappa r}}{\epsilon^2 r^6 (\ln{\frac{l_p}{a}})^2}   \\
E_{\perp} & \simeq & I \frac{l_p^2 a^4 e^{-2\kappa r}}{\epsilon^2 r^6}.
\end{eqnarray}
For $l_p \simeq 10$nm, $\epsilon \simeq 80$, $a \simeq r \simeq 1/2$nm, $I\simeq 1$eV, and $\kappa^{-1}\simeq 1$nm, we find that $E_{\parallel}\simeq 10^4 k_BT$ and $E_{\perp}\simeq 2 k_BT$.
At high salt concentrations, when the electrostatic screening length $\kappa^{-1} \simeq 2 \AA$, these dipolar interactions are suppressed by a factor of $\simeq 10^4$ suggesting that the binding mechanism discussed in the article dominates at short interchain separation.  At lower salt concentrations van der Waals interactions play a larger role in the formation of the aggregates, but the structure of those aggregates will still depend on the interchain electron tunneling mechanism.

\end{document}